\documentclass[a4paper,11pt,onecolumn]{article}
\usepackage{latexsym}
\usepackage{graphicx}

\title{\bf Catalog of averaged magnetic phase curves of stars }
\author{V.D. Bychkov$^{1}$, L.V. Bychkova$^{1}$, J. Madej$^{2}$  \\
$^{1}$ Special Astrophysical Observatory, Russian Academy of Sciences \\
$^{2}$ Warsaw University Observatory, Poland  }

\date{    }
\begin{document}

\begin{titlepage}
\maketitle

\begin{abstract}
The second version of the catalog contains information 
about 275 stars of different types. During the time that has 
elapsed since the creation of the first catalog, situation
fundamentally changed primarily due to the significant increase 
of accuracy of magnetic fields (MF) measurements. Up to now
global magnetic field were discoverd and measured in stars
of many types and their behavior partially was studied. Magnetic 
behavior of Ap and Bp stars is the most thoroughly studied. The catalog 
contains information about 182 such objects. The main goals for the
construction of the catalog are:
1. Review and summarize our kowledge about magnetic behavior of 
different types of stars.
2. The whole data are uniformly presented and processed which will 
allow one to perform statistical analysis of the variability of 
the longitudinal magnetic fields of stars.
3. The informations are presented in the most convenient form for
testing different theoretical models of different kind.
4. The catalog will be useful for the development of observational
programs.
\end{abstract}

\vspace{1.0cm}

  \begin{flushright}
    \vspace{2cm}
    {\it submitted to }  \\[2mm]
    {\rm Stars: from collapse to collapse } \\
    {\rm Special Astrophysical Observatory} \\
    {\rm 2016 October 3-7, Nizhnij Arkhyz} \\
    {\rm Karachai-Cherkessian Republic (Russia) } \\[15mm]
  \end{flushright} 
\par\vspace{3mm}

\end{titlepage}

\section{Introduction}

We present and analyze results of the longitudinal magnetic field $B_e$
measurements of stars collected from different bibliographic sources 
and personal communications, as well as our own determinations obtained 
at Special Astrophysical Observatory (Russian Academy of Sciences), see 
Bychkov (2008) and Bychkov et al. (2012).

Figs. 1-4 present sample rotational magnetic phase curves (MPC) for stars
of various types and the distribution of the available MPC over spectral 
classes.

\section{Description of the catalog }

For each star in the catalog the average magnetic phase curve was 
obtained by the least squares fitting of a sine wave or a double wave to
observed $B_e$ points. This is the same method as that used in the first
version of the catalog (Bychkov et al. 2005). In general the second 
version of the catalog is mostly very similar to the first version.

Recently the amount of available data sharply increased partly due to
the flow of new high-precision observational data obtained with the
Least Squares Deconvolution method (LSD) described by Donati et al.
(1997). These circumstances caused discovery of variable longitudinal
magnetic fields in a number of stars of different types and more
precise determination of MPC's in magnetic stars already known.

Fig. 1 shows the distribution of stars with known MPC among spectral types.
As the figure shows, the bulk of stars for which MPC's are known is 
concentrated in the spectral region around type A. This is due to the 
large contribution from Ap/Bp stars.

Table 1 shows the types of stars for which MPC's were determined. Precise
specification of types of stars sometimes is not possible. For example,
HD96446 consists of a He-r and $\beta $ Cep stars, HD97048 consists of
TTS (T Tau) and Ae/Be Herbig objects, flaring DT Vir consists of 
UV Cet + RS CVn types (Flare + RS CVn type stars) .

\begin{figure}[ht!]
\includegraphics[width=6cm]{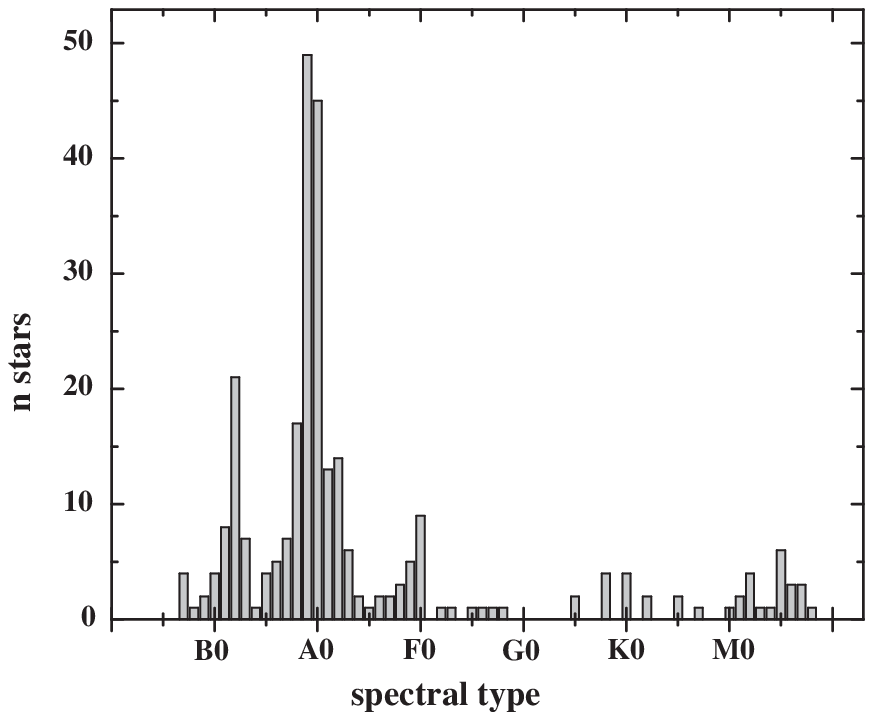}
\includegraphics[width=6cm]{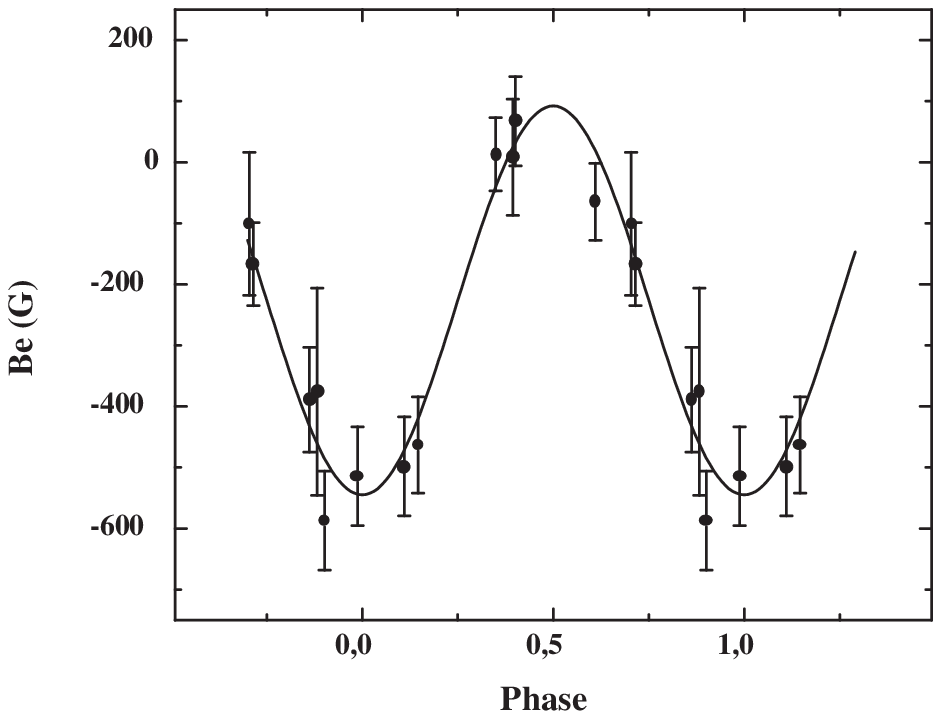}
\caption{ Left: Distribution of stars with known magnetic phase curves 
over spectral types. Right: Magnetic phase curve of the Ofp? star HD191612
with a rotational period of 537.6 days (Wade et al. 2011), the example
of supermassive supergiants of the Ofp? type. }
\label{fig:1}
\end{figure}

\begin{table}
\label{table:1}
\begin{center}
\caption{ Number of objects of various types with known MPC }
\vspace{2mm}
\begin{tabular}{|l|r|l|r|}
\hline
Ap/Bp                        & 182& TTS                          &   3 \\
var.beta Cep type            &  14& var.Ori type                 &   2 \\
Flare stars                  &  12& host Planet star             &   2 \\
multiple star                &  11& Be-star                      &   2 \\
Ae/Be                        &   9& semi-reguliar var.pulsating  &   2 \\
Rotationally var.star        &   7& Hosts a planetary system     &   2 \\
SPBS type                    &   6& var.delta Cep type           &   1 \\
HPMS                         &   5& Pulsating star               &   1 \\
var.BY Dra                   &   5& var.type delta Sct           &   1 \\
Of                           &   4& DA                           &   1 \\
normal chem.comp.stars       &   4& WN D                         &   1 \\
\hline
\end{tabular}
\end{center}
\end{table}

\begin{figure}[ht!]
\includegraphics[width=6cm]{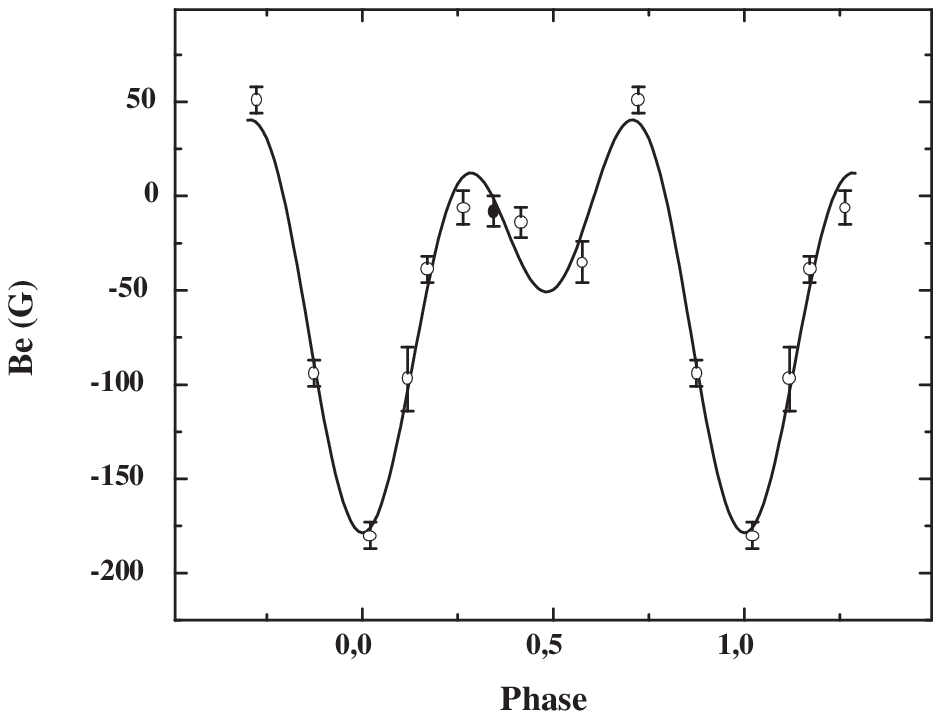}
\includegraphics[width=6cm]{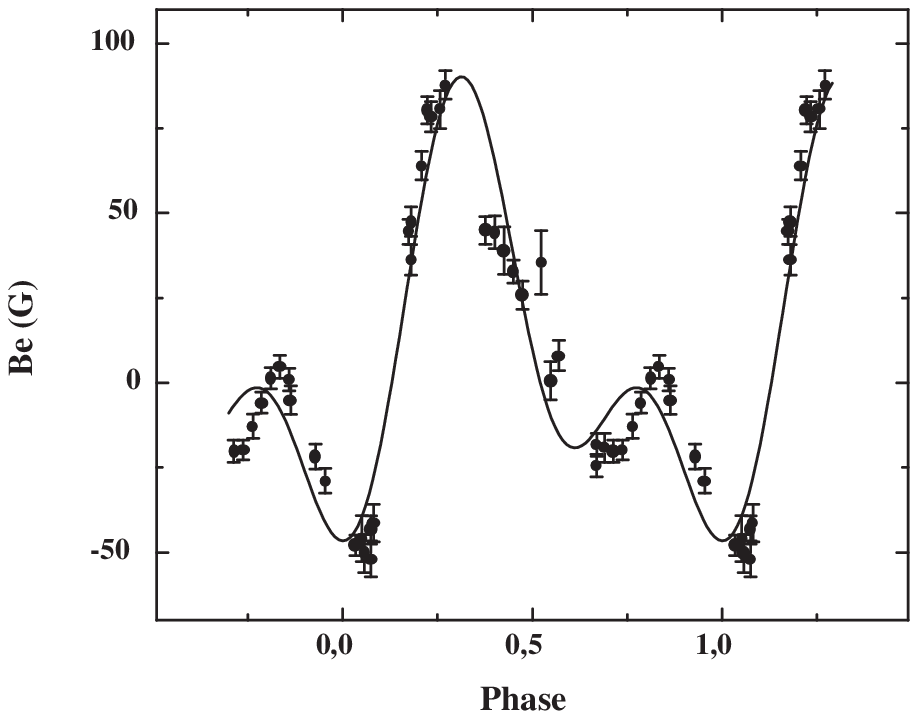}
\caption{Left:Classical T Tau star (CTTS) V2129 Oph Magnetic phase 
curve for the T-Tauri type star V2129 Oph. The period of rotation 
equals 6.53 days, according to Donati et al. (2007). Right:Magnetic 
phase curve for chemically normal stars HD149438, assuming the rotational
period of 41.033 days from Donati et al. (2006). }
\label{fig:2}
\end{figure}

\begin{figure}[ht!]
\includegraphics[width=6cm]{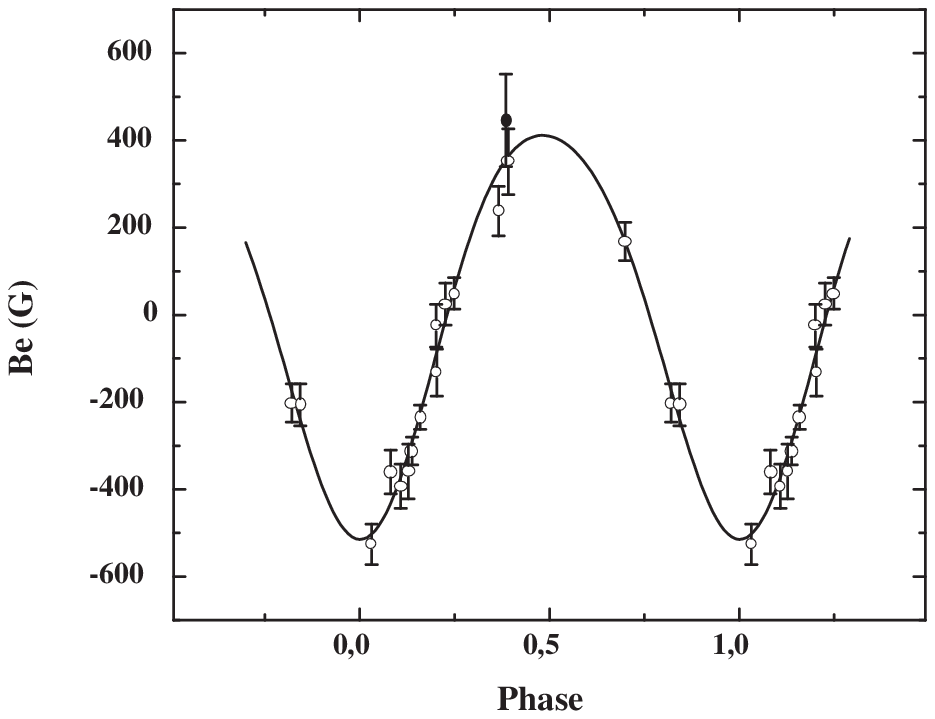}
\includegraphics[width=6cm]{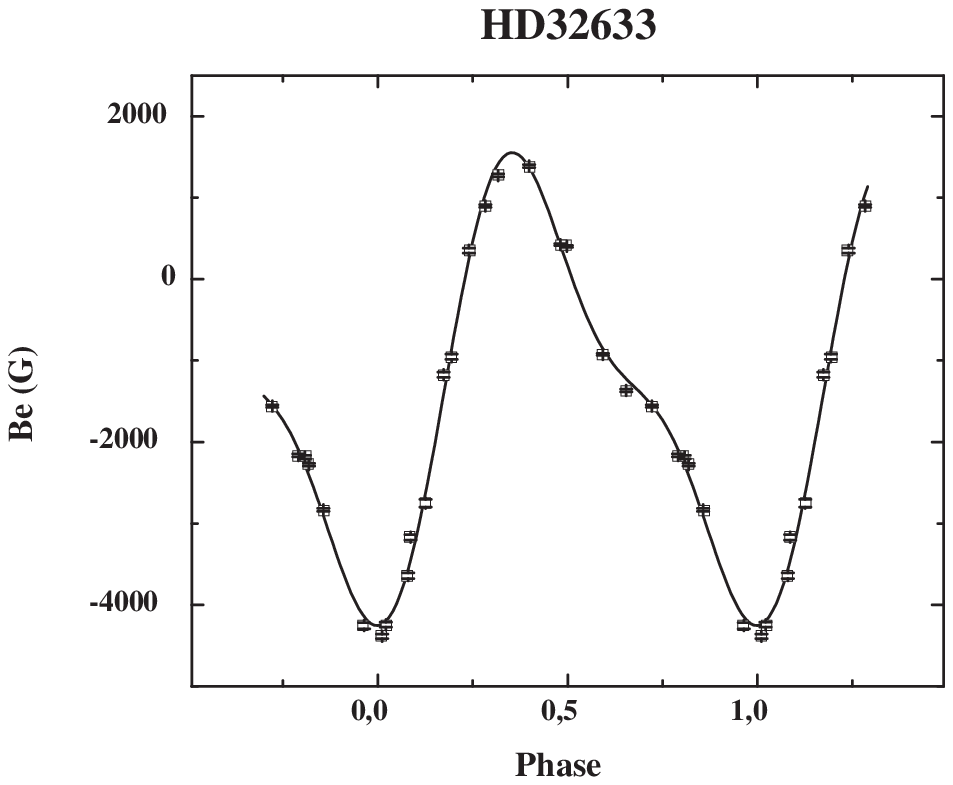}
\caption{Left:Magnetic phase curve of the Ae/Be Herbig star HD101412 with
the rotational period of 42.076 days by Hubrig et al. (2011), 
see also Jarvinen et al. (2016). Right:Magnetic rotational phase 
curve for the mCP star HD 32633 obtained from highly accurate observations
by Silvester et al. (2012). }
\label{fig:3}
\end{figure}

\begin{figure}[ht!]
\includegraphics[width=6cm]{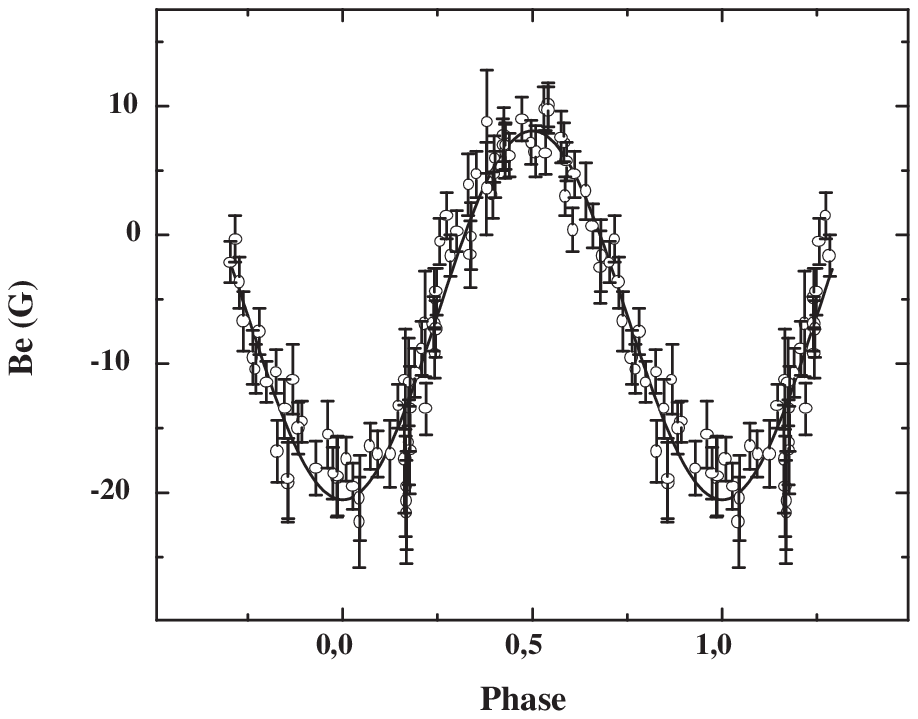}
\includegraphics[width=6cm]{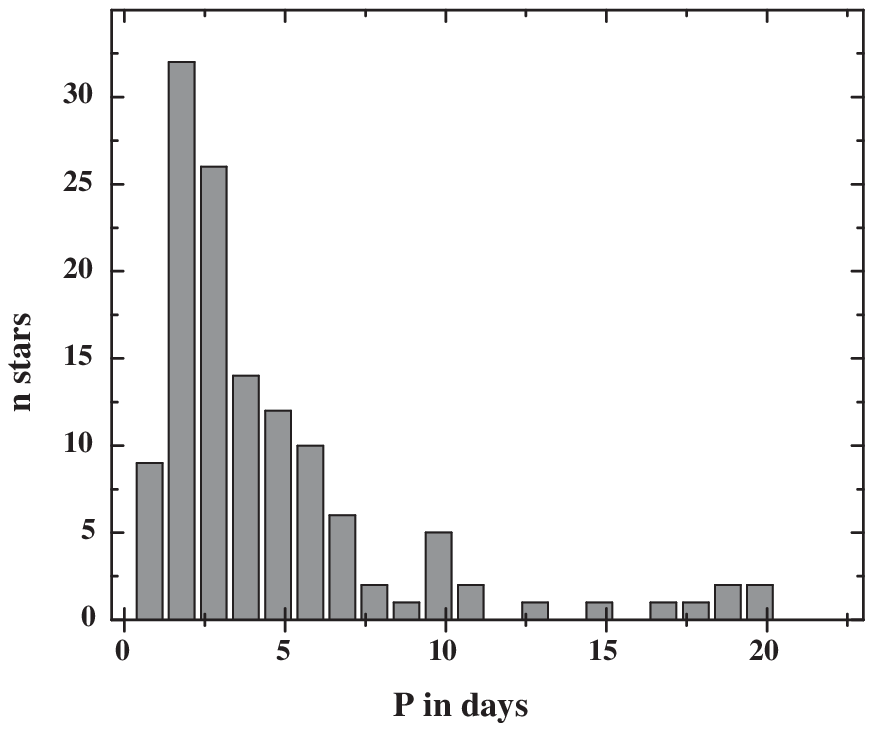}
\caption{ Left: Magnetic rotational phase curve for the SPBS star 
HD 3360 obtained from highly accurate observations by Briquet et al. (2016).
Right:Distribution of periods for Ap stars with simple harmonic 
magnetic phase curves. }
\label{fig:4}
\end{figure}

\begin{figure}[ht!]
\includegraphics[width=6cm]{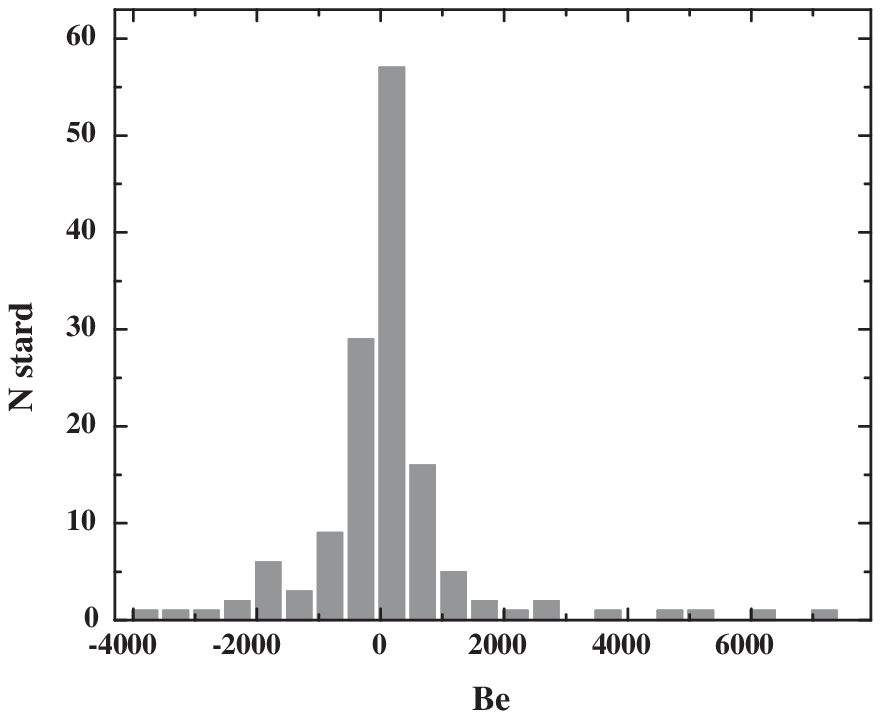}
\includegraphics[width=6cm]{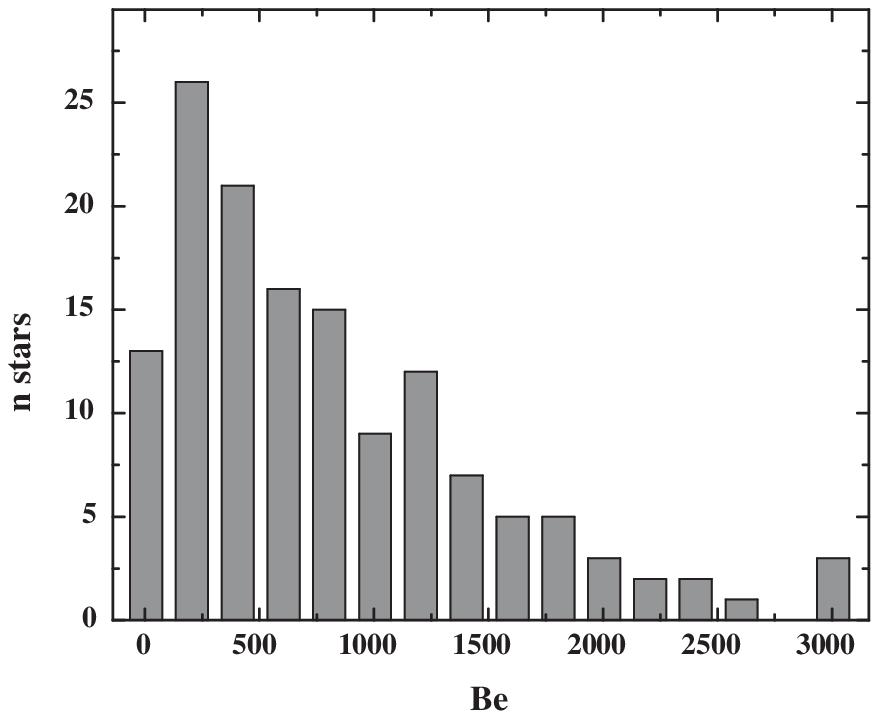}
\caption{Left: Distribution of the parameter $B_{0}$ for Ap stars with 
simple MPC. Right:Distribution of the halfamplitude $B_{1}$ for Ap stars
with simple MPC. }
\label{fig:5}
\end{figure}

\begin{figure}[ht!]
\includegraphics[width=6cm]{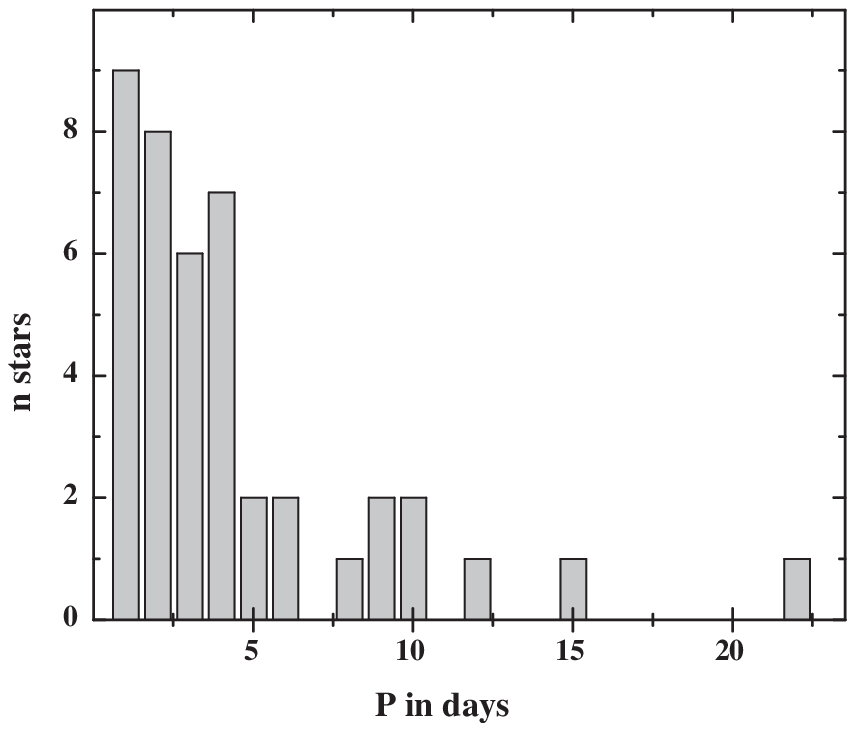}
\includegraphics[width=6cm]{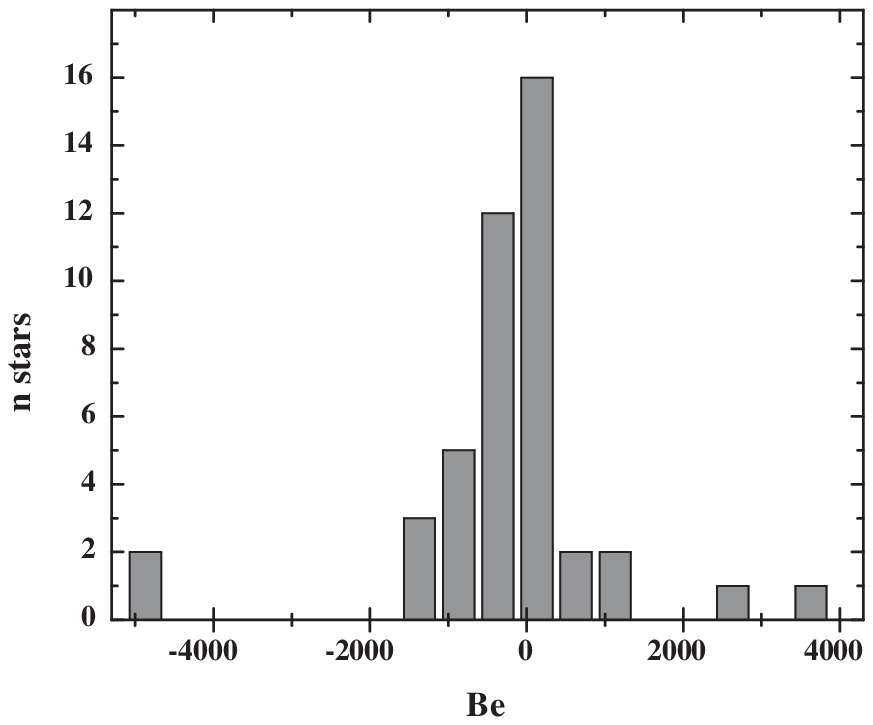}
\caption{ Left:Distribution of periods for Ap stars with complex MPC. 
Right:Distribution of the parameter $B_{0}$ for Ap stars with complex MPC. }
\label{fig:6}
\end{figure}

\begin{figure}[ht!]
\includegraphics[width=6cm]{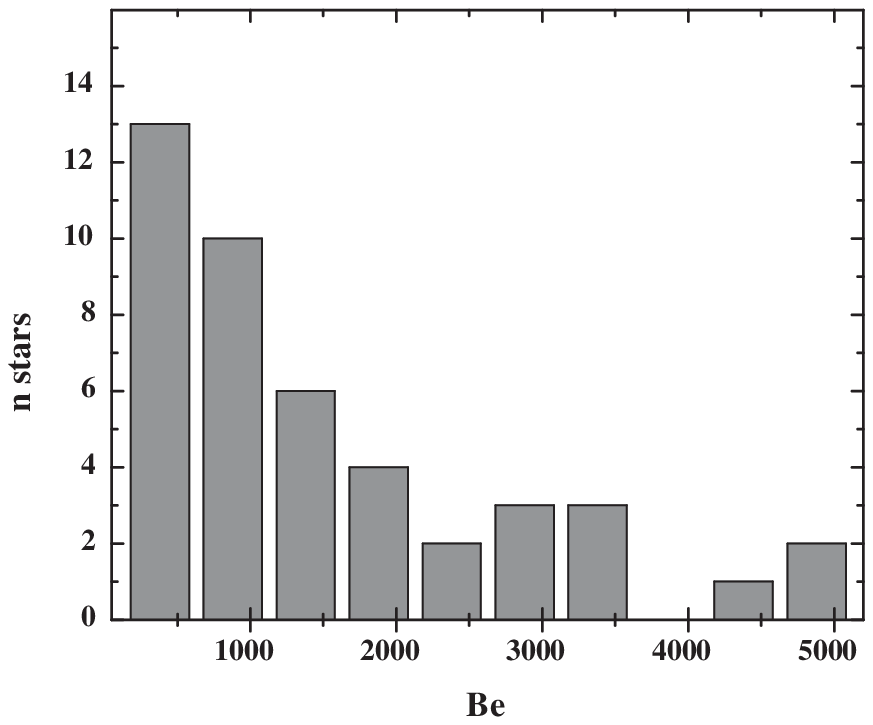}
\includegraphics[width=6cm]{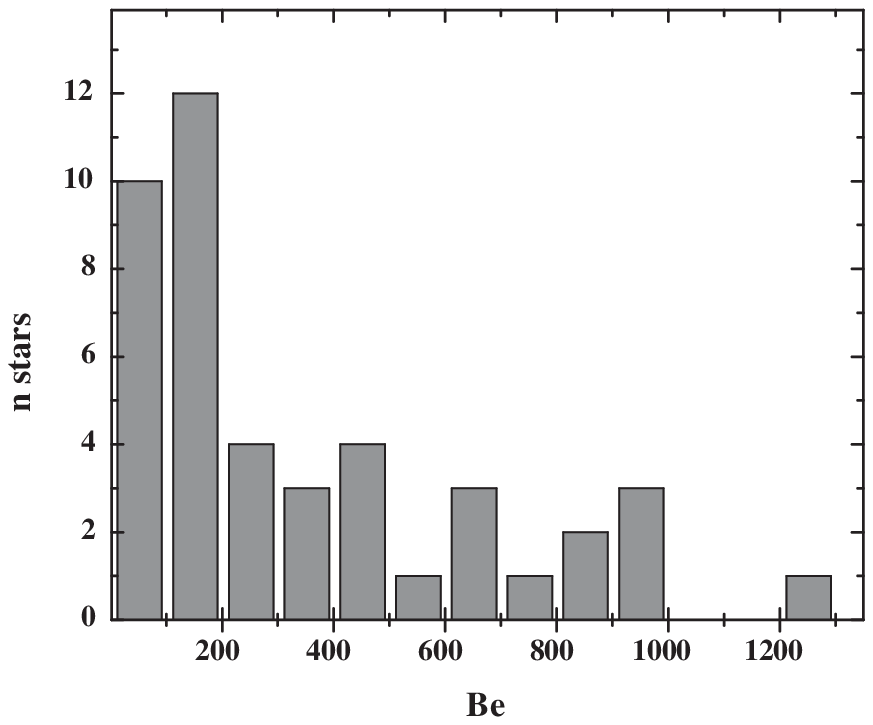}
\caption{Left:Distribution of the halfamplitude $B_{1}$ for Ap stars with complex MPC.
Right:Distribution of the halfamplitude $B_{2}$ for Ap stars with complex MPC.}
\label{fig:7}
\end{figure}

The best investigated stars in the Ap/Bp group, in turn, can be divided 
into two subclasses. The first subclass consists of stars who have simple 
sine wave MPC (total 141 objects). Fig. 4 shows the distribution of
rotational periods in this group of stars. Stars with long periods of 
rotation higher than the maximum period in Fig. 4 are: HD2453 period 
520.5 d, HD5797 -- 68.046 d, HD8441 -- 69.2 d, HD37058 -- period 223.75 d
HD41403 -- 4750. d, HD55719 -- 740.247 d, HD187474 -- 2345.8 d, HD188041
- 222.63 and HD201601 -- 35462.5 days.

Fig. 5 shows the distribution of coefficiens $B_0$ (average field
strength) and $B_1$ (half amplitude) of the best sine wave fits for 
stars with simple MPC.

The second group of Ap/Bp stars form objects whom IFC are double sine 
wave. This group consists of 45 stars. Fig. 6 shows the distribution 
of rotational periods in this group. Stars with extremely long periods
did not fit into Fig. 6: these are HD965 -- period 7125.65 d, HD9996 -- 
7961.8 d, HD18078 -- 1358. d, HD94660 -- 2800.0 d and HD126515 -- 
129.95 days.

Figs. 6-7 depict distribution of the coefficients $B_0$, $B_1$ and 
$B_2$ of double sine waves best fitted to the magnetic $B_e$ time series
observed in these stars. In the first version of the catalog (Bychkov 
et al. 2005) we pointed to the unusually large bias of $B_0$ towards
negative values. We attributed the effect to the low statistics of total
19 double wave stars in the old catalog. Actually we have 45 stars in 
the sample, over twice more than before. Now the distribution of $B_0$
is more symmetric around the value of zero and only a slight imbalance 
towards the negative values of the average $B_0$ still remains there.

Only in one exceptional star HD37776, the MPC is best described by 
the triple sinusoid.

\section{Conclusions}

Catalog of the magnetic rotational phase curves MPC primarily represents
a review and synthesis of the longitudinal magnetic field measurements
of stars of various classes, accumulated up to now. Principal reason of
the longitudinal field variability is rotation of a star with its magnetic
field frozen into ionized gas and seen at different aspect angles. 

Magnetic field itself evolves relatively rapidly, which can be observed
in cool red dwarfs (flare stars). This is a consequence of the strongly
differential rotation of those stars, which causes steady generation of 
local magnetic fields, which sometimes annihilate one another causing flares.
On the contrary, sometimes local fields add up to change strength and 
configuration of the global magnetic field in time scales from months to
years. This causes changes of parameters and shape of the magnetic phase curve.

In case of stars of other types, global magnetic fields do not evolve
rapidly. Nevertheless, there are apparent indications that in some stars
there are signatures of rapid evolution of the global magnetic fields.

We collected and uniformly processed all available observations on the
variability of longitudinal magnetic fields of stars in the unified form,
which allows one for a statistical analysis of the data and for testing and
verification of various theoretical models. The catalog will be also useful
for the development of observational programs. We believe, that the
accumulated observational data will be used to refine rotational periods
and parameters of many stars included in this catalog.

\section{Acknowledgements }
This work was supported by the Russian Science Foundation (project No.
14-50-00043).


\begin{thebibliography}{}

\bibitem{} Briquet, M., Neiner, C., Petit, P., Leroy, B., de Batz, B., 2016,
   A\&A, 587, A126

\bibitem{} Bychkov, V.D., Bychkova, L.V., Madej, J. 2005, A\&A, 430, 1143 
\bibitem{} Bychkov, V.D., Bychkova, L.V., Madej, J. 2016, in press

\bibitem{} Bychkov, V.D., Bychkova, L.V., Madej, J., Panferov, A.A. 2015, 
          ASP conf. ser., 494, 210

\bibitem{} Bychkov V.D., Bychkova L.V., Madej J., Topilskaya G.P. 2016,
          arXiv:1606.09562

\bibitem{} Donati, J.F., Semel, M., Carter, B.D., Rees, D.E., Cameron, A.C.
          1997, MNRAS, 291, 658

\bibitem{} Donati, J.-F., Howarth, I.D., Jardine, M.M., et al. 2006, MNRAS, 370, 629

\bibitem{} Donati, J.-F., Jardine, M.M., Gregory, S.G., et al. 2007, MNRAS, 380, 1297

\bibitem{} Donati, J.-F., Morin, J., Petit, P. et al. 2008, MNRAS, 390, 545

\bibitem{} Hubrig, S., Mikulasek, Z., Gonzalez, J.F., et al. 2011, A\&A, 525, L4

\bibitem{} Jarvinen, S.P., Hubrig, S., Scholler, M., Ilyin, I., Carroll, T.A.,
   Korhonen, H., 2016, AN, 337, 329

\bibitem{} Silvester, J., Wade, G.A., Kochukhov, O., et al. 2012, MNRAS, 426, 1003

\bibitem{} Wade, G.A., Bagnulo, S., Drouin, D., Landstreet, J.D., Monin, D.,
   2007, MNRAS, 376, 1145

\bibitem{} Wade, G.A., Howarth, I.D., Townsend, R.H.D., et al. 2011, MNRAS, 416, 3160

\end{thebibliography}
\end{document}